\documentclass[proceedings]{JHEP3} 
\PrHEP{AHEP2003}
\usepackage{epsfig}                   

\conference{International Workshop on Astroparticle and High Energy Physics}

\title{Search for neutrino radiative decays during a total solar eclipse}

\author{\speaker{Vlad Popa}\thanks{for the NOTTE Collaboration: S. 
Cecchini, D. Centomo, G. Giacomelli,
R. Giacomelli, V. Popa, C.G. \c{S}erb\u{a}nu\c{t} and R. Serra}\\
        INFN, I-40129 Bologna, Italy and ISS, R-77125 Bucharest, Romania\\
        E-mail: \email{popa@bo.infn.it}}

\abstract{We present the results of the measurements performed in the occasion of the 2001 
total solar eclipse, looking for visible photons emitted through a possible radiative decay
of solar neutrinos. We establish lower limits for the $\nu_2$ and $\nu_3$ proper lifetimes 
above $10^3$ s/eV, for neutrino masses larger than $10^{-2}$ eV.}

\begin{document}

  \section{Introduction}

The data accumulated in the last few years in favor of neutrino oscillations (both solar,
\cite{sk,sno1,sno2} and atmospheric \cite{sk,macro,miri}) demonstrate that neutrinos have
non vanishing masses (for recent reviews see \cite{gg,pak}). The most common interpretation
of the oscillation data is based on  two-generation mixing scenarios; the solar
$\nu_e$ neutrinos are supposed to be a mixing of two mass eigenstates, $\nu_1$ and $\nu_2$,
with $m_2 > m_1$ (in the ``normal hierarchy"):
\begin{equation}
| \nu_e \rangle = |\nu_1 \rangle \cos \theta_{12} + |\nu_2 \rangle \sin \theta_{12},
\label{mix}
\end{equation}
where $\theta_{12}$ is the mixing angle. A more complete description would require the
consideration of three neutrino generation mixing, but the available data do not allow
to determine all the corresponding mixing parameters.

If the neutrino mass states have also a non-vanishing magnetic moments, radiative decays
\begin{equation}
\nu_i \rightarrow \nu_j + \gamma
\label{decay}
\end{equation}
with $m_i > m_j$ could be possible, as initially hypothesized in \cite{sciama}; the first
 searches
for such decays were based on astrophysical considerations (see eg. \cite{cow}). The
status of the decaying theory and phenomenology was summarized in \cite{sciama2}.

The astrophysical neutrino lifetime lower limits are usually large (e.g. $\tau_0/m
> 2.8 \times 10^{15}$ s eV$^{-1}$ where $\tau_0$ is the lower proper lifetime
limit for a neutrino of mass $m$, \cite{blud}), but they are indirect and
rather speculative limits.

Much lower ``semi-indirect" limits were deduced from the re-interpretation of solar and
atmospheric neutrino data. Earlier attempts to explain the solar neutrino or
atmospheric neutrino anomalies only in terms of neutrino decay have been ruled out
by the existing evidence \cite{out}; the present accepted
explanations are based on neutrino oscillations, but
do not exclude the hypothesis of neutrino decays.
As an example,  from the SNO data \cite{sno1,sno2} a proper lower limit of
$\tau_0/m > 8.7 \times 10^{-5}$ s eV$^{-1}$ was deduced \cite{ab}. By analyzing
 all available solar neutrino data, other limits were obtained:
$\tau_0/m > 2.27 \times 10^{-5}$ s eV$^{-1}$ for the MSW solution,
and $\tau_0/m > 2.78 \times 10^{-5}$ s eV$^{-1}$ for the vacuum oscillation solution of
the
solar neutrino problem (SNP) \cite{aj}, or, following a different approach, $\tau_0/m
> 10^{-4}$ s eV$^{-1}$ \cite{jb}.

Direct searches for radiative neutrino decays have been also performed. As an example
we quote here the search for decay photons in the visible spectrum performed in
the vicinity of a nuclear reactor \cite{reactor}, yielding $\tau_0/m$ lower
limits in the range  $10^{-8}$ to nearly 0.1 s eV$^{-1}$, assuming neutrino
relative mass differences $\Delta m / m$ between 10$^{-7}$ and 0.1.
 Recently,
a search for $\gamma$ photons, using the Prototype
Borexino Detector at Gran Sasso \cite{borex} reported $\tau_0/m$ lower limits
of $1.5 \times 10^3$ s eV$^{-1}$ (assuming a polarization parameter $\alpha =
-1$ for the parent neutrino), $ 4.4 \times 10^3$ s eV$^{-1}$ (for $\alpha =0$) and
$9.7 \times 10^3$ s eV$^{-1}$ (for $\alpha =+1$).

Total solar eclipses (TSO) represent a particular opportunity to look for radiative
decays of solar neutrinos in the visible spectrum,
 during their flight from the Moon to the Earth, inside the
shadow cone produced by the Moon. The first experiment based on this idea was
performed in October 24, 1995 \cite{vanucci}, and a lower limit for the $\nu_2$
proper lifetime $\tau_0$ of about $10^2$ s was obtained, assuming neutrino masses of few eV
and $\Delta m^2_{21} = m^2_2 - m^2_1 \simeq 10^{-5}$ eV$^2$.

Some of us intended to perform measurements along this line during the 1999 TSE, in Romania:
two experiments were prepared, one airborne and one at mountain altitude, but the
weather conditions made the observations impossible \cite{n1}. We could only Analise
a video film recorded by a local television, obtaining $\nu_2$ lower lifetime limits
$1.8 \times 10^{-2} < \tau_0/m < 14.5$ s eV$^{-1}$ \cite{n2,n3,n4}.

In this paper we present the results obtained from the analysis of our 2001 TSE
observations.

\section{Experimental data}

The July 21, 2001 TSE was observed from a location near Lusaka (Zambia) (14$^o$56' lat. S,
28$^o$14' long. E, and 1200 m a.s.l.), at about 8 km from the line of centrality. We
used two instruments: a digital videocamera with an optical zoom $10 \times$ and
an additional $2 \times$ lens (which we will refer in the folowing as device ``A") and
a small Matsukov - Cassegrain telescope ($\phi$ = 90 mm, $f$ = 1250 mm) coupled to a
digital camera (device ``B"). The fields covered by a single pixel where about $10" \times
10"$ for images A and about $1.14" \times 1.14"$ for B. The experimental data consist
in 4149 frames obtained with the videocamera (data set ``A") and 10 digital pictures
obtained with the telescope (data set ``B"). Fig. \ref{foto} shows two of those images.
\FIGURE[t]{\epsfig{file=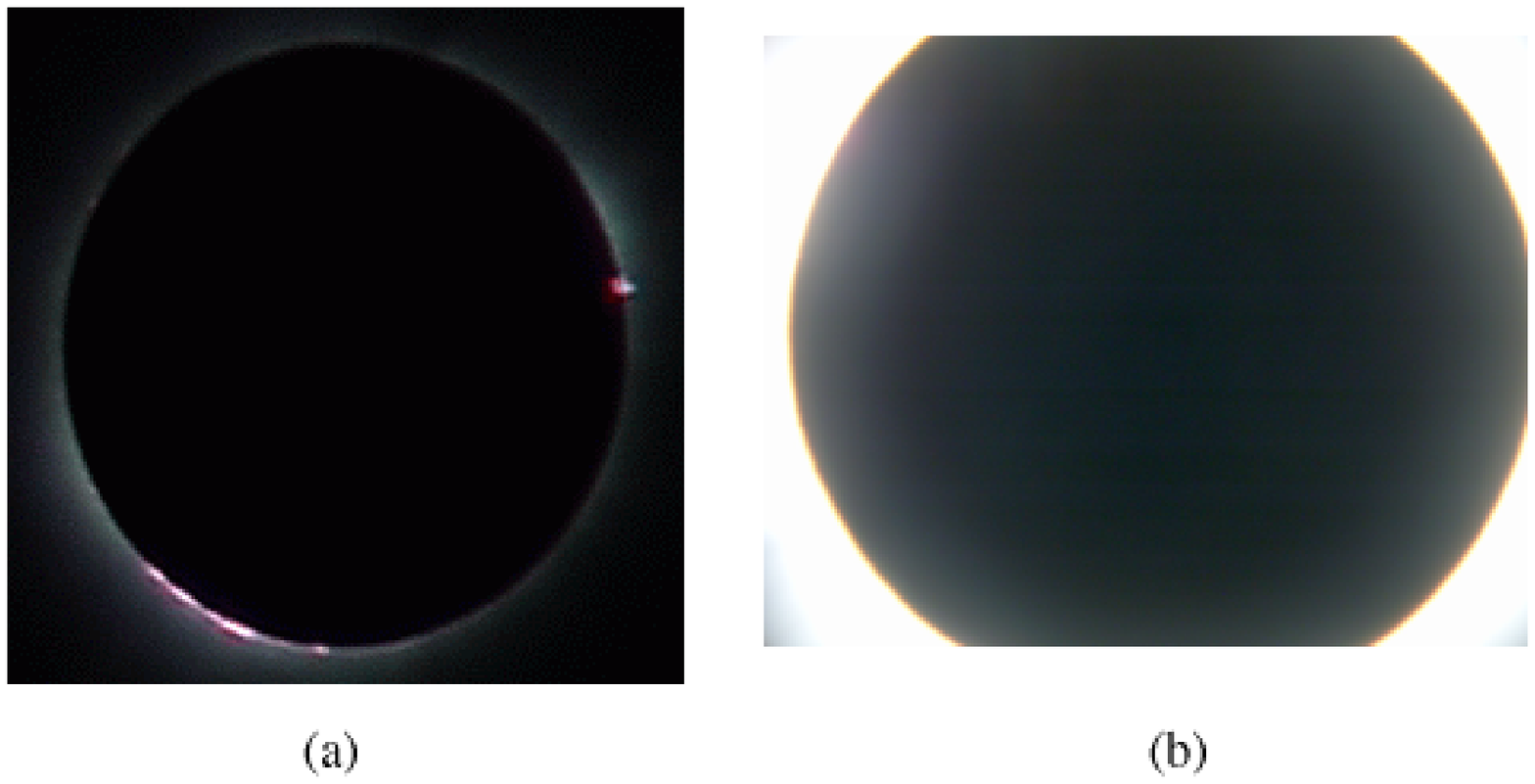, width=.8\textwidth}
	     \caption{Two sample images of the 2001 TSE extracted from (a) data set A, and
(b) data set B. Note that the pictures are not at the same scale.}\label{foto}}
For our analysis we summed the three color channels (Red, Green and Blue) of the images,
thus recomposing the ``white" light signal.
The totality phase of the TSE in our observation site was about 3.5 minutes, so the
displacement of the Sun behind the Moon is not negligible. As the signal of solar
neutrino decays should be correlated with the direction to the center of the Sun, we
calculated, for each frame in both data sets, the relative position of the Sun and, after
determining by fit the center of the Moon disk, we computed the shift to be
considered when summing the images in order to have the center of the Sun in the same pixel.

Both instruments were calibrated at the Catania and Bologna Astronomical Observatories.
In the exposure conditions of the eclipse, the number of visible photons required for
producing 1 ADU (Acquisition Digital Unit) was 7.3 10$^4$ for instrument A and 8.9 10$^2$
for device B.

\section{The simulation}

In order to extract physical information concerning a possible neutrino radiative decays
from TSE data, a previous knowledge of the characteristics of the expected signal in
mandatory. For the analysis of the 2001 data we developed a full 3-dimensional Monte
Carlo (MC) simulation \cite{nou}, based on the predictions of a recent version
of the Standard Solar Model \cite{bahcall}.
We first randomly chose a solar neutrino production reaction, and, consequently, a
neutrino energy and the point of its creation inside the core of the Sun.
Since we are interested only in neutrino decays that may produce signals in our
detectors, we then generate a random photon arrival direction, inside the angular
acceptance of our devices, and a decay point, uniformly distributed along the
photon path, between the observation point (the Earth) and the Moon.

This procedure does not respect the kinematic probabilities of the simulated decay,
so we attribute to each MC event a weight according the decay angular probability
density:
\begin{equation}
\frac{d\Gamma}{d \cos \theta^*} \propto \frac{m_i}{\left(\Delta m^2_{ij} \right)^3}
\left(m^2_i + m^2_j + m_i m_j \right) \left( 1+ \alpha \cos \theta^* \right).
\label{gamma}
\end{equation}
In Eq. \ref{gamma} $m_i$ and $m_j$ are the masses of the parent and respectively
daughter neutrino (see Eq. \ref{decay}), $\Delta m^2_{i,j} = m^2_i - m^2_j$, and
$\theta^*$ is the angle between the photon momentum and the spin of the initial
neutrino, in its center of mass reference frame. The polarization parameter $\alpha$
varies from -1 (left-handed) and 1 (right handed) for Dirac neutrinos, and is 0 for
Majorana neutrinos. The kinematic weight for each MC event is obtained by
integrating Eq. \ref{gamma}
over all photon directions that would lead to a signal in the same pixel, thus
considering the experimental angular resolution. Note that such weights are dependent
on the simulated device.

This MC incorporates both the realistic geometry of solar neutrino production and decay,
(as in \cite{frere}, where considering monoenergetic solar neutrinos the authors
obtained an analytical prediction) and the standard energy spectrum predicted by the SSM, (as in
our previous code \cite{n1,n2} where the neutrino source was approximated as pointlike).
We assumed that
$m_1 < m_2 < m_3$ where $m_1$, $m_2$, $m_3$ are the masses
of the $\nu_1$, $\nu_2$ and
$\nu_3$ mass eigenstates, respectively.  We
restrict our analysis to a two generation mixing
scenario, assuming the present
mass differences obtained from solar neutrino experiments,
the LMA solution with $\Delta m^2_{12} = 6 \times 10^{-5}$ eV$^2$.
Since SNO suggests also
the presence of $\nu_3$ in the solar neutrino flux, we considered also the mass
difference measured by atmospheric neutrino experiments:
$\Delta m^2_{13} \simeq
\Delta m^2_{23} = 2.5 \times 10^{-3}$ eV$^2$.

Fig. \ref{mcu} shows the expected ``luminosity curves" (the average luminosity versus
the angular distance from the center of the Sun) from the simulation of (a) $\nu_2
\rightarrow \nu_1 + \gamma$ and (b) $\nu_3 \rightarrow \nu_1 + \gamma$ decays. The
weights in Eq. \ref{gamma} where calculated for different $m_1$ values.
\FIGURE[t]{\epsfig{file=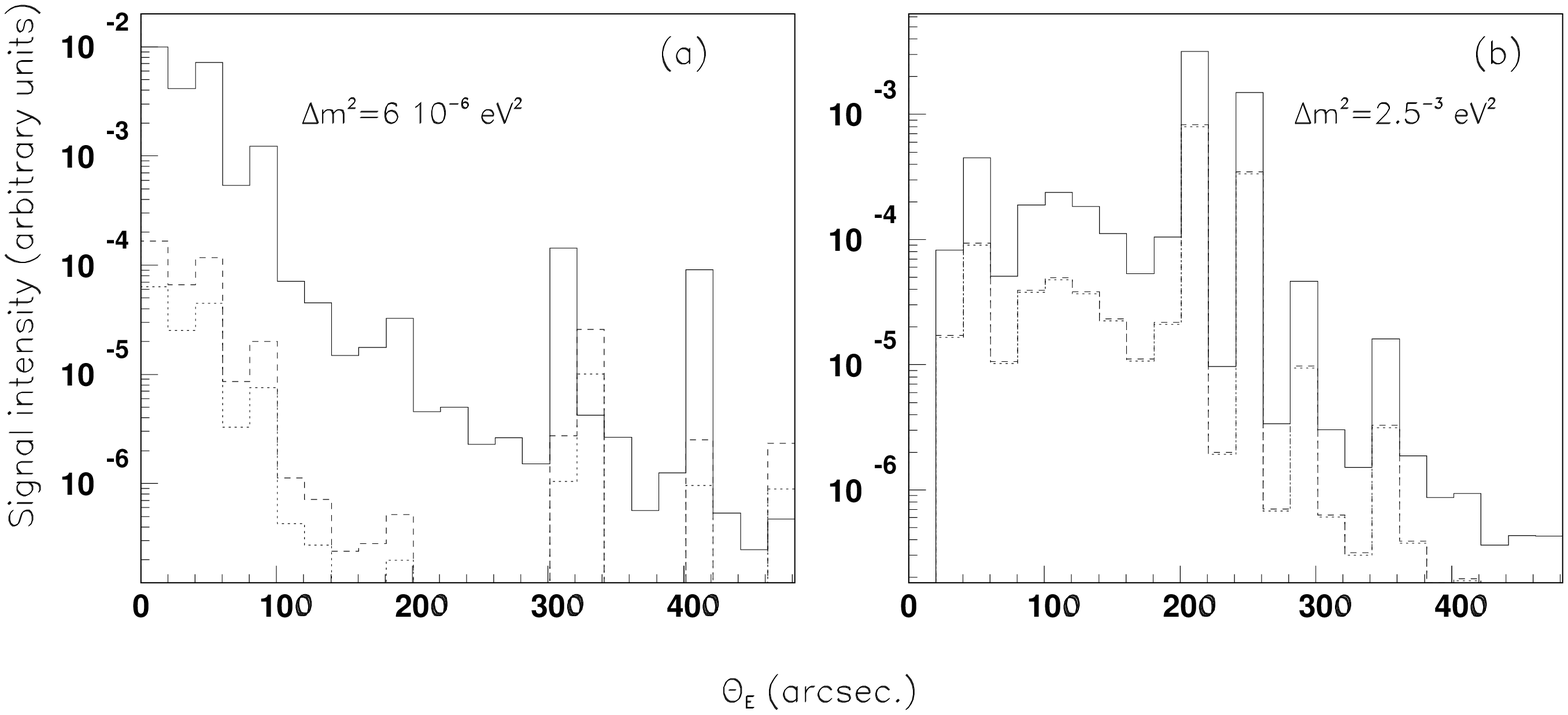, width=.9\textwidth}
\caption{The expected shapes of the visible signals produced by the
hypothesized solar neutrino radiative decay,
 assuming $m_1 = 0.001$ eV (solid histograms), 0.01 eV
(dashed) histograms) and 0.1 eV (dotted histograms). The squared mass difference
is assumed to be $6 \times 10^{-5}$ eV$^2$ (a) and $2.5 \times 10^{-3}$ eV$^2$ (b).
In all cases $\alpha = -1$.}\label{mcu}}
The histograms in Fig. \ref{mcu}a suggest that for all neutrino masses, 
the expected signal is concentrated at
small $\theta_E$ angles (about 50 arcsec).
The widths and shapes of the signals are
sensitive to the mass assumed: the larger the mass, the narrower the signal band.
In the case of Fig. \ref{mcu}b,
the signal is
broader (about 250 arcsec) and is less sensitive to the mass choice.

One of the main results of the MC simulation consists in the determination of the 
global probabilities $P$ of a solar neutrino  decay according
to Eq. \ref{decay}, during its flight from the Moon to the Earth, and,
 produces a visible photon that reaches the detector. Those
probabilities are both neutrino mass and instrument dependent.
Consequently,
assuming that an experiment detects $N_\gamma$ photons produced by neutrino radiative
decays, the lifetime of the neutrino can be calculated from
\begin{equation}
N_\gamma = P \Phi_i S_M t_{obs} \left(1-e^{-\frac{\langle t_{ME} \rangle}{\tau}}
 \right)
e^{-\frac{t_{SM}}{\tau}}~,
\label{prob}
\end{equation}
where
 $P$ are the probabilities estimated by the MC simulation,
  $\Phi_i = \Phi_\nu \sin^2 \theta_{1i}$, ($\Phi_\nu$ is the
flux of solar neutrinos at the Earth (or Moon) and $\theta_{1i}$ the mixing angle)
is the local flux of solar $\nu_i$ mass eigenstate neutrinos,
$S_M$ is the area of the
Moon surface covered by the analysis  and $t_{obs}$ is the time of observation.
$\langle t_{ME} \rangle$ is the average
time spent by solar neutrinos inside the observation cone
(one third of the flight
time from the Moon to the Earth), and $t_{SM}$ is the time of flight of the neutrinos
from the Sun to the Moon. 

A complete discussion of the MC and of its results may be found in \cite{nou}.

\section{Data analysis}

From the simulation described in the previous Section it follows that the expected visible
signal from solar neutrino decays would have specific angular scales. A proper tool for
investigating different scales of the eclipse images is the wavelet analysis. This 
technique gradually removes the contributions from various background sources as the 
diffraction of the coronal light on the borders of the Moon, the diffuse sky light, the 
ashen light (light reflected by the Earth on the surface of the Moon), etc.
 We used the
simple Haar wavelet basis \cite{wave}.
The $n$-order term of the decomposition is obtained by dividing the $N \times N$
pixels$^2$ image in square fields of $N/2^n \times N/2^n$ pixels$^2$ and averaging the
luminosity in each field; the averages are then removed and the resulting image,
the $n$-order residual, can be used to obtain the $(n+1)$-order term. Thus, each
decomposition term results in an image in which objects of the corresponding
scale are dominant, while the residuals contain  information for
smaller dimension scales.

As the wavelet analysis
requires a dyadic dimension of the field (the number of pixels on each border
of the image is a power of 2), we retained from our images $64 \times 64$ pixels
(from data A) and $512 \times 512$ pixels (from data B) defined around 
the position of the center of the Sun, and summed them in order to increase the
signal to background ratios.

The decay signal is searched for by averaging the luminosity of the images over
``rings" centered on the position of the center of the Sun.
There is no  ``central pixel"; so we have
considered each of the four pixels adjacent to the image center as ``central" and
then averaged the obtained luminosity profiles.
Such profiles, obtained from the raw data (before the wavelet decomposition) are shown in
Fig. \ref{raw}.
\FIGURE[t]{\epsfig{file=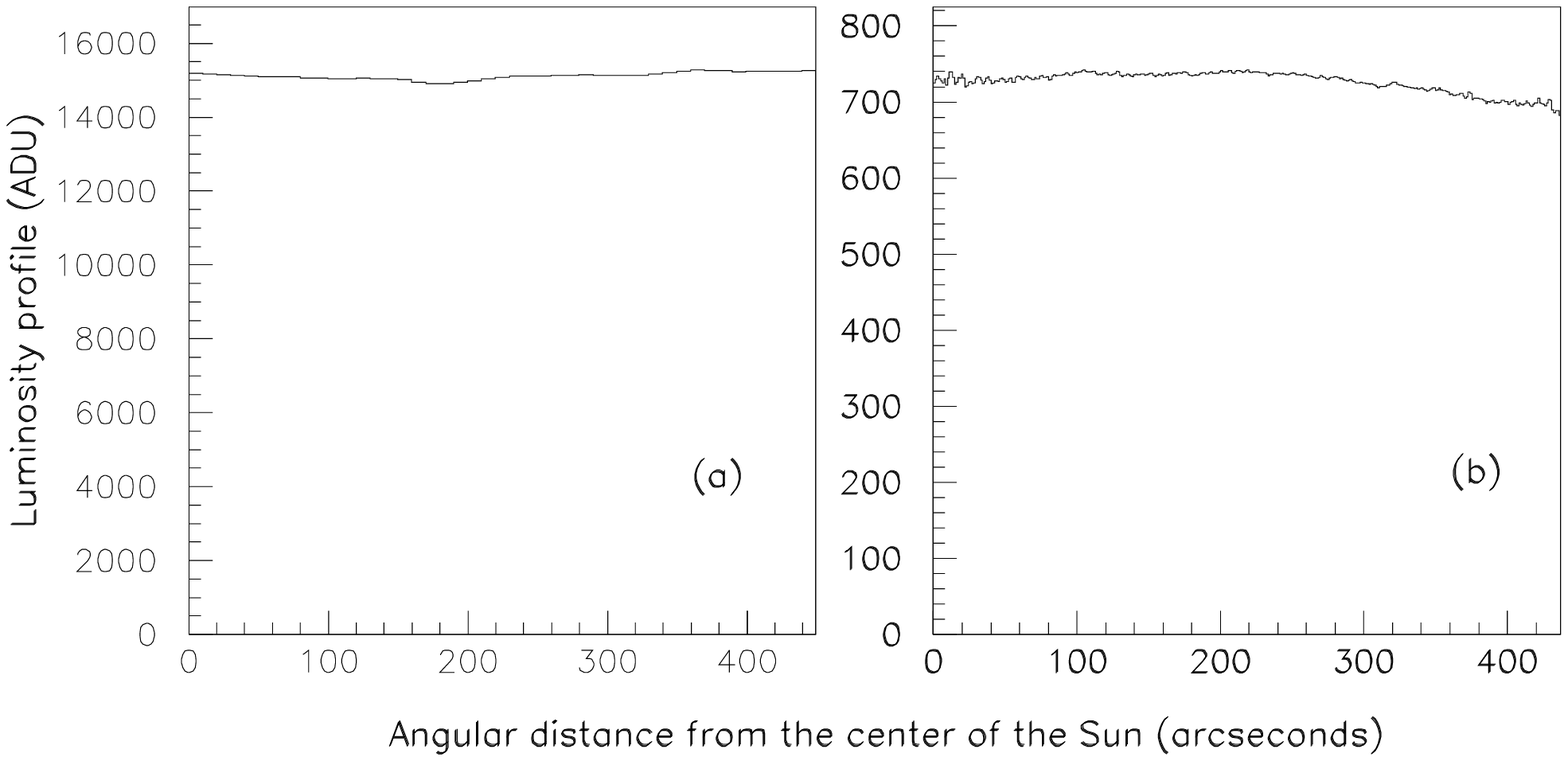, width=.9\textwidth}
\caption{The luminosity profiles obtained from the raw data: a) data set A, b)
data set B.}\label{raw}}
The  difference between the two total images is due to the different CCD
sensitivities, spatial resolution and optical features of the instruments.
Data set A presents
no clear structure, data B might contain some at relatively large $\theta$.
In order to check that the shape of the luminosity curve in data set B could be produced
by the ashen light, we aligned the 10 digital images along the direction of the 
center of the Moon, and made a similar analysis on a full Moon pictured obtained 
with the same instrument. Figs. \ref{figl} show this comparison. In order to enhance 
the contrast of the images, from both pictures we removed the average luminosity, 
thus obtaining the $0^{th}$-order residual of the wavelet decomposition. 
\FIGURE[t]{\epsfig{file=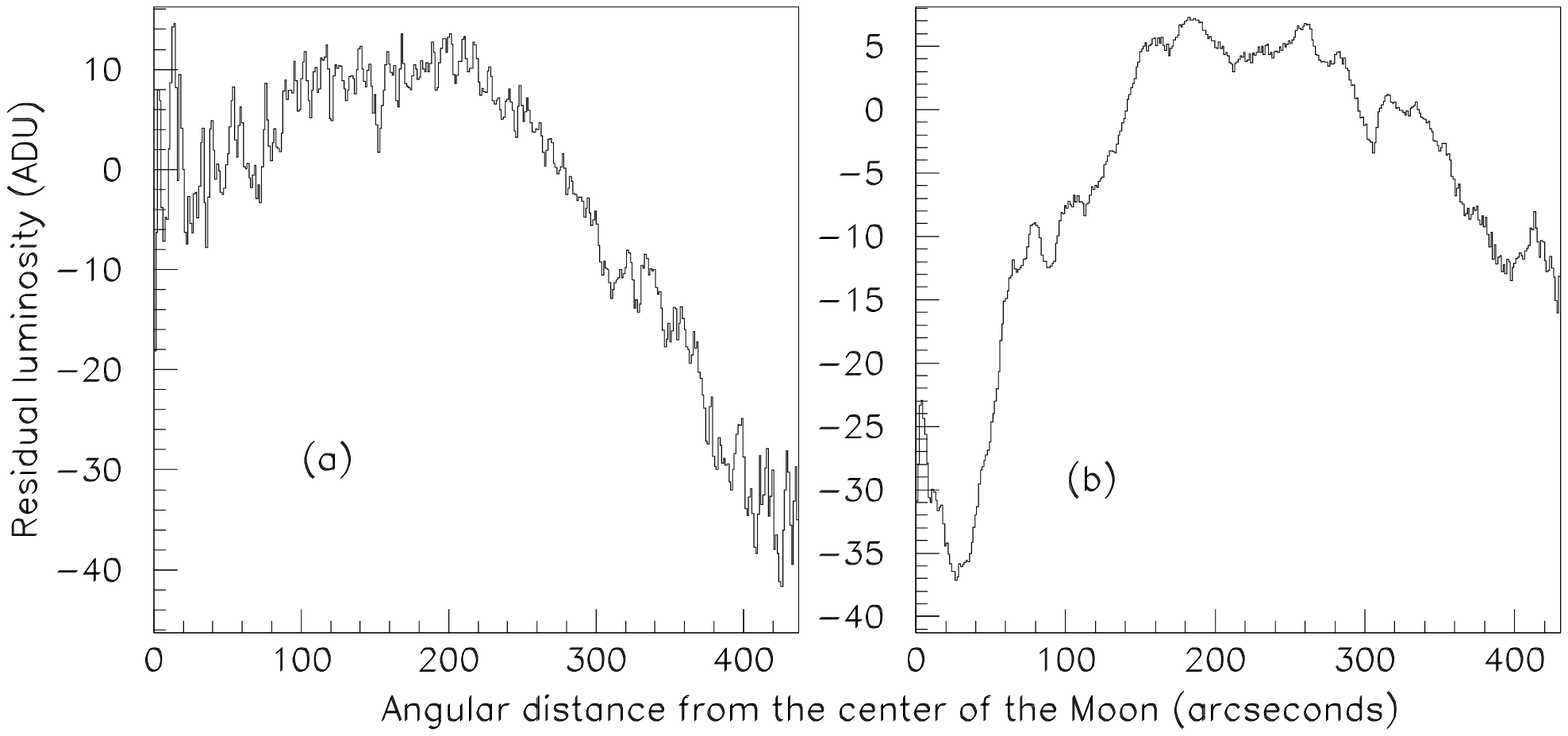, width=.9\textwidth}
\caption{Light luminosity profiles after removing the average luminosity of
(a) the sum of images in data set B aligned with respect to the center of the
Moon, and (b): an image of the full Moon obtained with the same instrument.}\label{figl}}

The structures in Fig. \ref{figl}a are  similar to those in Fig \ref{figl}b;  we should take into
consideration that the
Earth reflects the light of the Sun as a convex mirror, thus the central part of the Moon
receives more light from the Earth than the rest of it
(the relative excess in the raw TSE data is only about 2\%). Instead the Sun illuminates the
Moon uniformly. This observation suggests that we cannot simply remove the image of the
full Moon from the data, as it would create a fake signal in the central part of the
resulting image. Thus, one should develop a reliable model of the ashen
light; alternatively one should use the wavelet decomposition.
 Note  that the
exposure conditions for the image of the full Moon were different than those during
the eclipse, so the ADU values are not directly comparable.
As we cannot determine which is the real contribution of the ashen light in set B, we
can use the results only to determine a lower limit for the $\nu_3$ lifetime.

\subsection{Search for the $\nu_2 \rightarrow \nu_1 + \gamma$ signal}

The expected signal from a $\nu_2 \rightarrow \nu_1 + \gamma$ decay, considering
its MC estimated width \cite{nou},
should be better seen in the fourth order wavelet term
of data sets A and B, as it corresponds to structures with about 40" - 60" width.
Figs. \ref{figt4} show the luminosity distributions for this term. Note that
each bin is an average over $4 \times 4$ pixels in the case of data set A (Fig. 5a), and
over $32 \times 32$ pixels for set B (Fig. 5b).
\FIGURE[t]{\epsfig{file=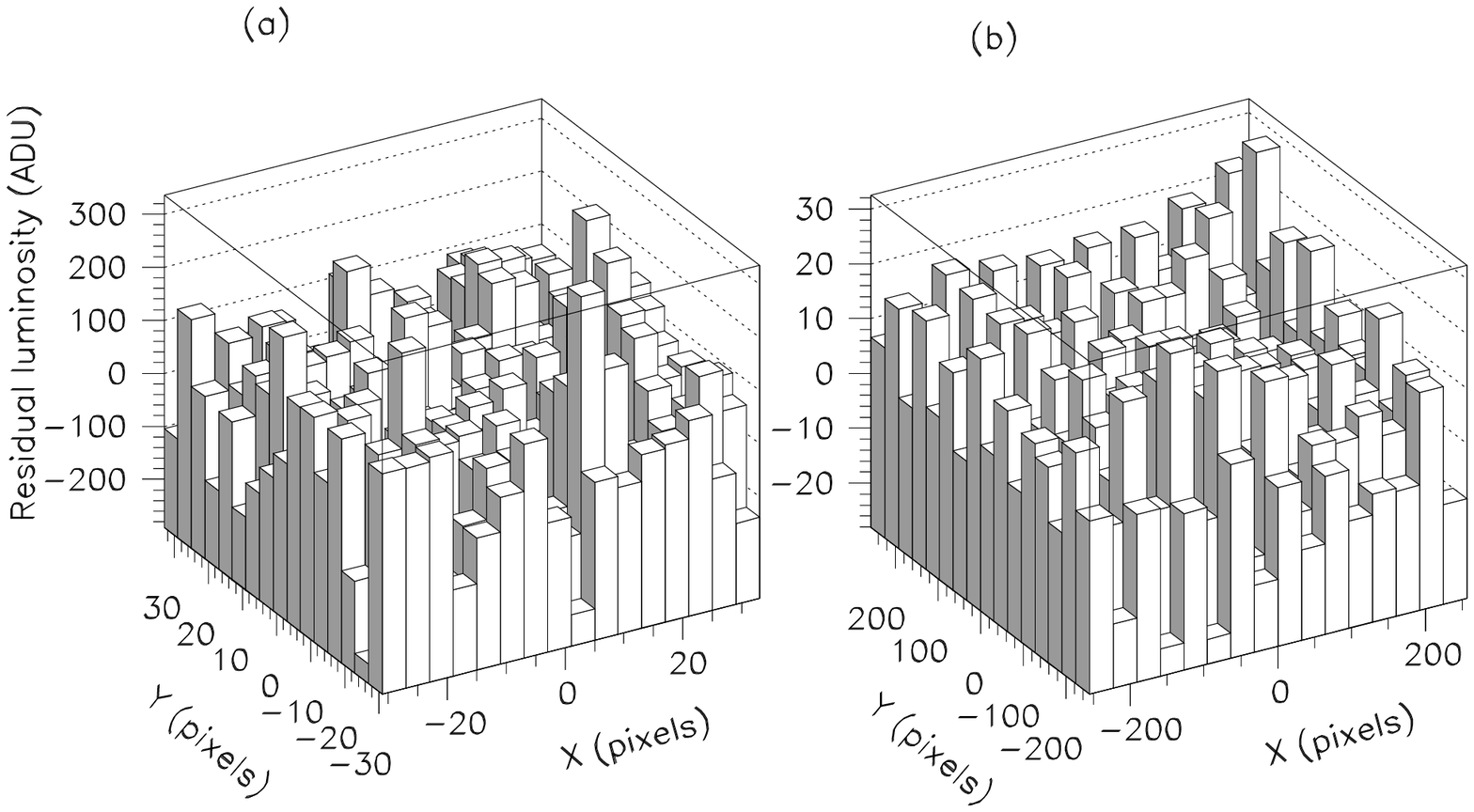, width=.9\textwidth}
\caption{``White" light luminosity distributions of the fourth order wavelet term of (a)
the summed images
A and (b) B  (centered on the Sun).}\label{figt4}}
 No central maximum is present in both data sets, and we can use these wavelet terms
to determine lower lifetime limits for the investigated decay.

The 95\% CL lower limits for the $\nu_2$ lifetime, in our case of no signal,
are obtained by the substituting in Eq. \ref{prob}
$N_\gamma$ with $3\sigma_{N_\gamma}$ of the forth order wavelet terms decomposition
of the data, and considering $\sin^2 \theta_{ij} = sin^2 \theta_{12} \simeq 0.74$
(the LMA solution of the ``Solar Neutrino Problem", \cite{sk,sno1,sno2}).
 They are shown with thicker lines
in Figs. \ref{lim2}a (data  A) and \ref{lim2}b (data  B),
assuming that $\nu_2$ is a Dirac (lefthanded or righthanded) or a Majorana neutrino.
The recent limits obtained from the  Borexino
 Counting Test Facility \cite{borex} are also shown, for comparison.
The arrows labelled ``SNO" and ``WMAP" indicate the lower neutrino mass limit
reported by SNO \cite{sno1,sno2}, and the upper mass limit obtained by WMAP \cite{map}.
The limit obtained by the first TSE experiment \cite{vanucci} is indicated by the
horizontal arrow; note that this limit,  obtained using different physical
hypotheses, is valid for neutrino masses of few eV.
\FIGURE[t]{\epsfig{file=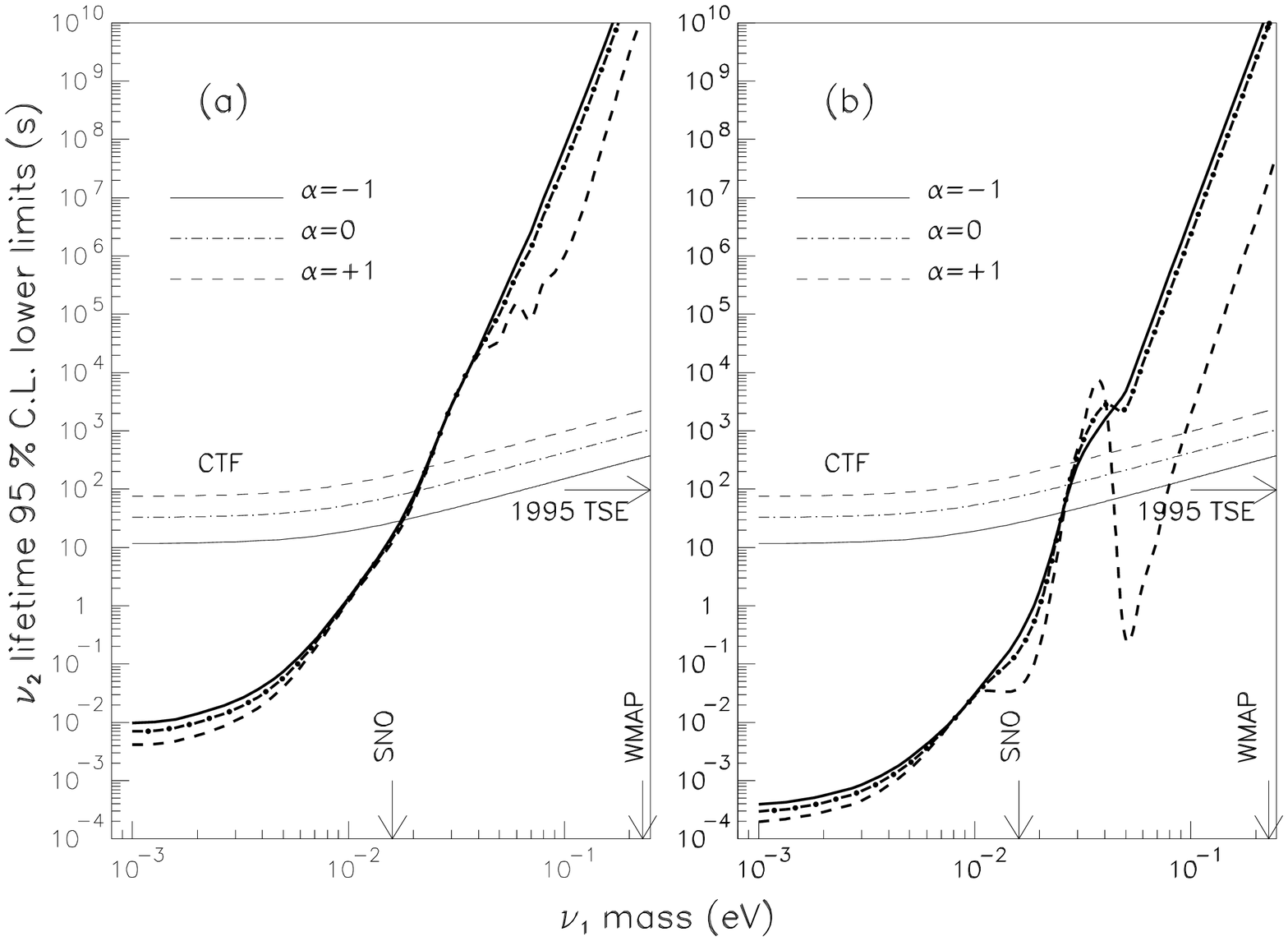, width=.9\textwidth}
\caption{The 95\% CL lower limits for the $\nu_2$ proper lifetime,
as function of the $m_{\nu_1}$, obtained from data sets (a) A and (b) B.
The results are valid in a neutrino mixing scenario with only two
generations, and $\Delta m^2_{2,1} = 6 \times 10^{-5}$ eV$^2$.
The discontinuities in the proper lifetime limits for righthanded neutrinos originate in
the MC probabilities and reflect the changements in the initial neutrino energy imposed by
the condition of obtaining visible decay photons pointing to the Earth.
Other relevant limits are also
indicated (see text).}\label{lim2}}

 Neutrino lifetime values larger than our lower limits
 are not in conflict with the oscillation explanation of the
solar neutrino deficit. The neutrino time of flight from the Sun to the Earth is
about 500 s (in the laboratory frame of reference). The Lorentz boost for a solar
neutrino with a mass of about 0.02 eV is $\gamma \simeq 1.5 \times 10^{7}$,
so the fraction of $\nu_2$ that would decay into $\nu_1 + \gamma$, assuming
$\tau_0 \simeq 60$s (in the c.m.) would be only  $\simeq 5 \times 10^{-7}$.

\subsection{Limits on the $\nu_3 \rightarrow \nu_{1,2} + \gamma$ lifetimes}

As already shown, the search for $\nu_3 \rightarrow \nu_{1,2} + \gamma$ signals is more
difficult, as the ashen light could create a fake signature. Furthermore, the expected
angular width of the expected signal is larger, so the wavelet decomposition could
erase it. We still can compute 95\% CL lower limits for the corresponding lifetimes,
substituting $N_\gamma$ in Eq. \ref{prob} with $3\sigma_{N_\gamma}$ of the raw data A and B.
The mixing angle $\theta_{13}$ is not known, but it should be small; we assume
$\sin^2 \theta_{13} \simeq 0.1$. The lifetime limits obtained in those conditions
are shown in Fig. \ref{lim3}.
\FIGURE[t]{\epsfig{file=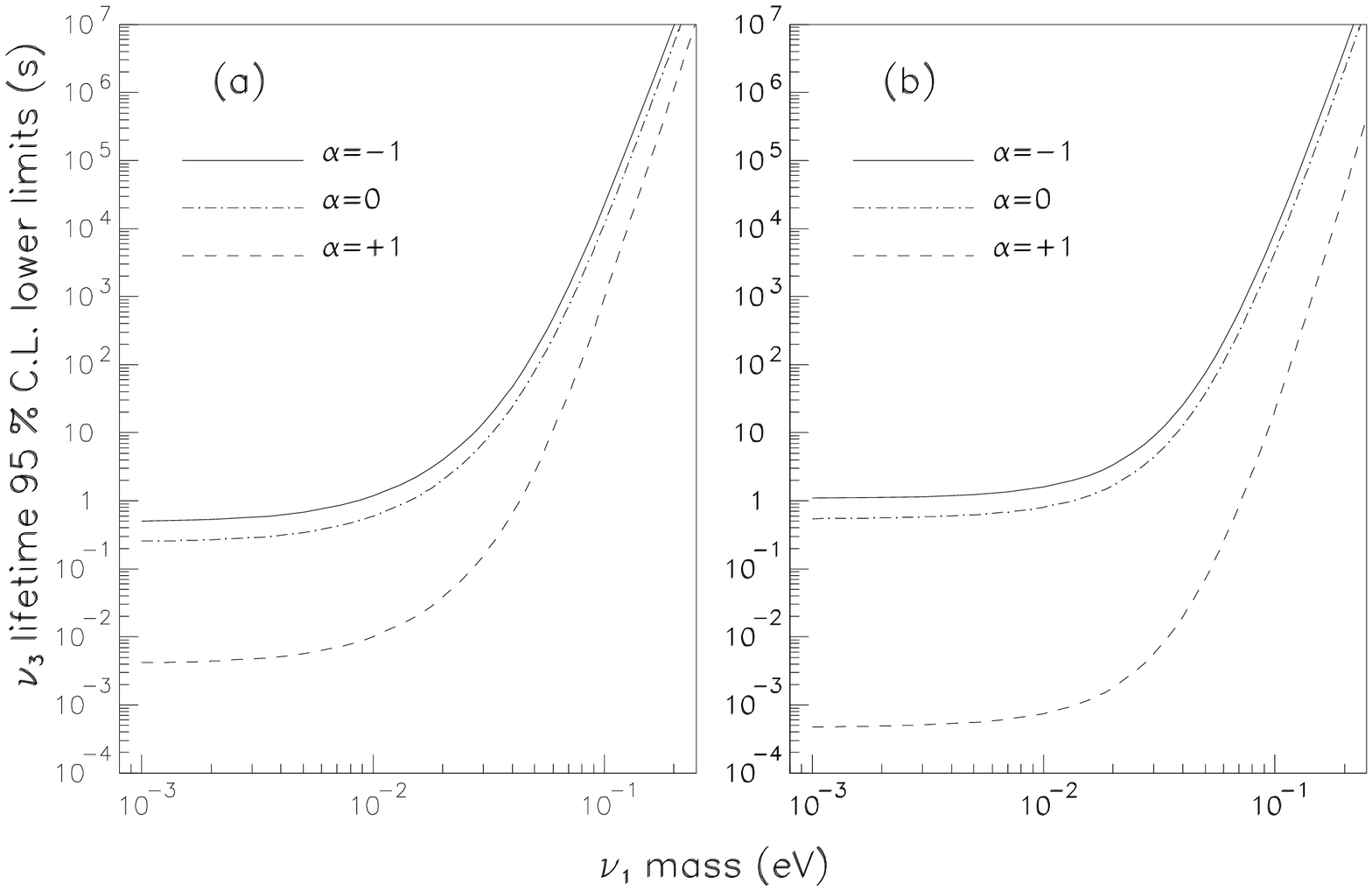, width=.9\textwidth}
\caption{The 95\% CL lower limits for the $\nu_3$ proper lifetime,
as function of the $\nu_1$ mass, obtained from (a) set A (a) and (b) B.
 The solid, dot-dashed and dashed lines
correspond to three different neutrino polarizations, $\alpha = 1$, 0 and -1,
respectively. These results are obtained assuming $\Delta m^2_{3,(2,1)}
= 2.5 \times 10^{-3}$ eV$^2$ and $\sin^2 \theta_{13} \simeq 0.1$}\label{lim3}}

\section{Conclusions}

We analyzed two sets of digital images obtained during the June 21$^{st}$ 2001
total solar
eclipse, in Zambia, looking for possible radiative decays of solar neutrinos, yielding
visible photons.

Data set A consists in a large number of frames  recorded
with a digital videocamera; it has a relatively large integration time, but
 a modest  space resolution.

Set B consists of 10  pictures taken with a
digital camera coupled to a small
telescope. Its time coverage is poorer than for set A,
but it has a better space resolution and
the instrument sensitivity was an order of magnitude better.

The proper lower lifetime limits
(95\% CL) obtained for the $\nu_2 \rightarrow \nu_1 + \gamma$
decays of lefthanded neutrinos range from
$\tau_0/m_2 \simeq 10$ s eV$^{-1}$ to  $\simeq
10^9$ s eV$^{-1}$, for $10^{-3}$ eV $< m_{\nu_1}
< 0.1$ eV, see Fig. \ref{lim2}.
 These limits are among the best
obtained from direct measurements, demonstrating the potentiality of
neutrino decay experiments during total solar eclipses (or possibly made in space, using
the Earth as light absorber \cite{frere}). The lab. lifetime limits
are about $10^7$ times
larger, thus the fraction of neutrino decays from the Sun to the Earth
would be negligible.

A similar analysis was made for a possible $\nu_3 \rightarrow \nu_{2,1} + \gamma$ decay,
assuming $\sin^{2} \theta_{31} \simeq 0.1$ (the  value of
this mixing angle is not known).
 No signal compatible with a possible $\nu_3 \rightarrow \nu_{2,1} + \gamma$
is
seen.
The obtained 95\% C.L. $\nu_3$ proper lifetime lower limits, for $m_1 \geq 10^{-2}$ eV and
for $\alpha = -1$, 0,
are about two orders of magnitude
lower than for the $\nu_2$, Fig. \ref{lim3}

 New observations,
 in better technical conditions,
during  forthcoming TSE's should be considered.

An attempt along these lines 
 was made during the December 2002 eclipse, but the weather conditions
in South Africa did not allow any observation. We
intended to use three portable telescopes, equipped with astronomy type CCD's. The
sensitivity would have been
about two orders of magnitude better than what reported
in  this paper.

\section{Acknowledgments}

We would like to acknowledge many colleagues for useful comments and discussions.
We thank the
people of the Catania and Bologna Astronomical Observatories for their assistance during
calibrations.
Warm thanks are due to the Kiboko Safari, Lilongwe, Malawi, for their assistance
during the expedition in Zambia.
This work was funded by NATO Grant PST.CLG.977691 and partially supported by
the Italian Space Agency (ASI), INFN and the Romanian Space Agency (ROSA).
V.P. thanks the organizers of the AHEP-2003 Workshop, Valencia, for their hospitality.

\end{document}